\documentclass{article}
\usepackage{graphicx}
\def\p{\partial}
\def\s{\sigma}

\def\g{\gamma}

\def\d{\delta}
\def\de{\delta}
\def\De{\Delta}

\def\ld{\lambda}
\def\Ld{\Lambda}
\def\L{\Lambda}

\def\e{\eta}

\def\Om{\Omega}
\def\rh{\rho}

\def\b{\beta}

\def\a{\alpha}

\def\pdellx'{\frac{\partial}{\partial x'}}
\def\pdellw'{\frac{\partial}{\partial w'}}
\newcommand{\be}{\begin{equation}}
\newcommand{\ee}{\end{equation}}
\def\bed{\begin{displaymath}}
\def\eed{\end{displaymath}}
\def\bea{\begin{eqnarray}}
\def\eea{\end{eqncrray}}
\def\[{$$}
\def\]{$$}
\begin{document}
\title{Space-time translational gauge identities\\  
in Abelian Yang-Mills gravity  }
%\vspace{0.2in}
%\bigskip

\author{ Jong-Ping Hsu\footnote{e-mail: jhsu@umassd.edu}\\
Department of Physics,
 University of Massachusetts Dartmouth \\
 North Dartmouth, MA 02747-2300, USA}

 %Unlike a word processing program such as Microsoft's {\em Word\/} or
%Sun's
%{\Large \bf   Appendix B.}
\bigskip
%\end{center}
\maketitle
{\small  We derive and calculate the space-time translational gauge identities in quantum Yang-Mills gravity with a general class of gauge conditions involving two arbitrary parameters.  These identities of the Abelian group of translation are a generalization of  Ward-Takahasi-Fradkin identities and important for general discussions of possible renormalization of Yang-Mills gravity with translational gauge symmetry.  The gauge identities in Yang-Mills gravity with a general class of gauge conditions are 
substantiated by explicit calculations.}
%{\Large \bf    Total Unified Model of All Interactions }
% \makeatletter
 %\renewcommand\theequation{{13.}\@arabic\c@equation} % for (1)
  %\renewcommand\thesection{{13-}\@arabic\c@section} 
%\renewcommand\theequation{\thechapter.\@arabic\c@equation} % for (1.1)
%\makeatother
 %\noindent
%\bigskip
%******************
%{\em OpenOffice\/}, \LaTeX\ is a mark-up language and a typesetting
%program.
%\section{Introduction\noindent
 
\section{Introduction}

In quantum electrodynamics, the Abelian $U_1$ gauge symmetry implies that a linear gauge condition can be imposed at all times and that there are no interacting ghosts.\cite{1}  This local gauge symmetry leads to the conservation of the electric charge and the Ward-Takahasi-Fradkin (WTF) identities\cite{2,3,4}, which imply relations between different renormalization constants.  In non-Abelian gauge theories with internal gauge groups, such as $SU_2$ theory, a gauge condition cannot be imposed at all times.\cite{5}  As a result, there emerges a new Lagrangian involving additional couplings of gauge fields and anti-commuting scalar fields,  which are are called Faddeev-Popov ghosts and appear only in the intermediate steps of a physical process.  Thus, the non-Abelian gauge symmetry leads to a generalization of WTF identities, which are called Slavnov-Taylor identities.\cite{6,7}  The gauge groups in electroweak theory and chromodynamics are all internal and compact Lie groups.

In contrast, Yang-Mills gravity is based on an external translational gauge group ($T_4$) in flat space-time, which is an Abelian subgroup of the Poincar\'e group and is a non-compact Lie group.  It has both the Abelian $T_4$ gauge symmetry and 
non-linear gauge field equations, so that a gauge condition for the $T_4$ Abelian gauge field  cannot be imposed at all times.  In order to restore gauge invariance and unitarity of Yang-Mills gravity, we must have additional coupling of $T_4$ gauge fields $\phi_{\mu\nu}$  and anti-commuting vector fields $V_\mu (x)$ and $V'_\nu (x)$,\cite{8} just like non-Abelian gauge theory.  It was demonstrated that Yang-Mills gravity in flat space-time is consistent with all classical tests of gravity, including gravitational radiations.\cite{9,10} Such an experimental consistency depends on a crucial property. Namely, an `effective Riemann metric tensor' emerges in and only in the geometric-optics limit of the photon and matter wave equations.\cite{9,11}  Furthermore, we can construct a unified model for Yang-Mills gravity, electroweak theory and chromodynamics,\cite{12,13}  following the unification idea of Glashow-Weinberg-Salam.

Yang-Mills gravity is based on the symmetry of the following transformation in flat space-time,
\be
x'^\mu = x^\mu + \Ld^{\mu}(x),
\ee
%%1
where $\Ld^{\mu}(x)$ is an infinitesimal arbitrary gauge function.  It is important to note that the infinitesimal transformations (1) has dual interpretations\cite{9}:  (i)  They are the local flat space-time translations, and (ii)  they are also arbitrary coordinate transformations.  In this sense, the dual interpretation of (1) implies that the local translational gauge symmetry is equivalent to the general flat space-time symmetry.  The physical meaning of the space-time transformations (1) implies [a] translations, [b] rotations, and [c] distortions of coordinates (which are associated with space-time transformations of non-inertial frames.)\cite{14}   The generators of the space-time translation $p_{\mu}$ has the representation $+ i\p/\p x^{\mu} \ (c= \hbar=1)$.
  Thus, symmetric tensor fields $\phi_{\mu\nu}$ naturally appear in the $T_4$ covariant gauge derivative\cite{15}
  \be
  \De_{\mu} =\p_{\mu} -i g \phi_{\mu}^{\nu} p_{\nu} =  \p_{\mu} + g\phi_{\mu}^{\nu}\p_{\nu} \equiv J_{\mu}^{\nu}\p_{\nu},
  \ee
  %%%%%%%%%2
  $$
J_{\mu}^{\nu} = \de_{\mu}^{\nu} + g \phi_{\mu}^{\nu},
$$
through the interesting combination of  $ \de_{\mu}^{\nu}$ and $g \phi_{\mu}^{\nu}$ in the tensor $J_{\mu}^{\nu}$.  We have seen that the tensor field $\phi_{\mu\nu}$ is associated with the most general flat space-time symmetry, as shown in the transformations (1).  Thus, the gravitational tensor  field $\phi_{\mu\nu}$ may be termed as the space-time translational gauge field or  the space-time gauge field.  The absence of $i=\sqrt{-1}$ in the $T_4$ covariant gauge derivative (2) implies the impossibility for the presence of both attractive and repulsive force in the interaction of the tensor gauge field.  Furthermore, the coupling constant $g$ of the space-time gauge field has the dimension of length, in sharp contrast to the dimensionless coupling  constants in electroweak theory and quantum chromodynamics.  These properties based on the space-time symmetry in the transformations (1) and its dual interpretation are just right for gravity.\cite{8,9}  

Associated with the coordinate transformations (1), the $T_4$ gauge transformations for fields can be defined.\cite{9} For example, we have
 \be
T_{\mu\nu} \to (T_{\mu\nu})^{\$}=T_{\mu\nu} - \L^\ld \p_\ld T_{\mu\nu} -
T_{\mu\a} \p_\nu
\L^\a - T_{\a\nu} \p_\mu \L^\a, 
 \ee
%%%%%%%13.7%%%3
\be
Q^{\mu\nu} \to (Q^{\mu\nu})^{\$}= Q^{\mu\nu} - \L^\ld \p_\ld Q^{\mu\nu} +
Q^{\ld\nu} \p_\ld \L^\mu + Q^{\mu\ld} \p_\ld \L^\nu, \ etc.
\ee
%%%%%%%%%4
In the derivation of the $T_4$ gauge identities, we impose as usual the following constraint for the gauge function $\Ld_\ld $
\be
\Ld_{\ld} = [U^{-1} (J)]_{\ld\s} \rho^\s,
\ee
%13.9%%%%%%5
where $\rho^\s$ is an arbitrarily infinitesimal function independent of $\phi_{\mu\nu}(x)$ and the operator $U$ is given by equation (14) below.

Equations (1) - (4) describe the space-time symmetry of Yang-Mills gravity in flat space-time.  In general, the $T_4$ gauge theory can be formulated in arbitrary coordinates, which can accommodate both inertial and non-inertial frames.\cite{9} In this paper, we shall use inertial frames with the metric tensor $\e_{\mu\nu}=(1,-1,-1,-1)$ for discussions.   One should not say in general that the space-time gauge symmetry dictates the existence of the tensor field because a free tensor field theory can also be invariant under the gauge transformations (3) and (4).  This is in contrast to usual gauge theories with internal gauge groups, where one could say that gauge symmetry dictates the existence of interacting gauge fields.  However, it seems fair to say that one of the simplest tensor fields with a universal coupling to all matter is associated with the  translational gauge symmetry in flat space-time and with the gauge covariant derivative (2).

\section{Space-time gauge identities}

In Yang-Mills gravity, the difficulties involved in calculations, especially in the higher orders, are formidable, even though the interaction structure is much more simple than that of Einstein gravity in curved space-time.  Fortunately, we also have space-time gauge identities which can be used to double check on the consistency of calculations.  Gauge identities can relate different renormalizable constants and are important for general discussions of possible `renormalization' of a theory.

We shall derive the space-time gauge identities for a wide class of gauge conditions involving two gauge parameters $\xi$ and $\zeta$ in Yang-Mills gravity by using the general method of Slavnov.\cite{6}
 Let us consider the following class of gauge conditions
\be
F_\ld (J) = Y_{\ld} (x), \ \ \ \ \ \    J_{\mu\nu} = \e_{\mu\nu} + g\phi_{\mu\nu}
\ee 
%%%%%%%%%13.1%%%%6
\be
 F_{\ld} (J) \equiv \frac{1}{2}(\de_{\ld}^{\mu} \p^{\nu} + \de_{\ld}^{\nu} \p^{\mu} - \zeta \e^{\mu\nu}\p_{\ld}) 
 J_{\mu\nu} \equiv  h_{\ld}^{ \  \mu\nu} J_{\mu\nu},
\ee
%%13.2 %%%%7
where $Y^{\nu}$ is independent of 
the fields and the gauge function $\L^{\mu}$. 
Similar to that in the usual gauge-invariant theory,\cite{16,8}
the  vacuum-to-vacuum amplitudes with the gauge condition (6) is given by the path integral,
\be
W(Y^{\ld})= \int d[\phi]
exp(iS_{\phi})
 \ det \ U  \ \mbox{\large \boldmath $ \Pi$}_{x,\ld}\d( F_{\ld}(J) -Y_{\ld}),
\ee
%%8
where $d[\phi] \equiv d[\phi_{\a\b}] $ and the gravitational action $S_{\phi}$ is given by 
\be
S_{\phi} = \int d^{4}x \ L_{\phi}, \ \ \ \  L_{\phi}= \frac{1}{4g^2}\left (C_{\mu\nu\a}C^{\mu\nu\a}- 
2C_{\mu\a}^{ \ \ \  \a}C^{\mu\b}_{ \ \ \  \b} \right), 
\ee
%%%%%%%%%%%%2%%11%%%9
\be
C^{\mu\nu\a}= J^{\mu\s}\p_{\s} J^{\nu\a}-J^{\nu\s} 
\p_{\s} J^{\mu\a}, \ \ \ \ \  J_{\mu\nu}=\eta_{\mu\nu}+ 
g \phi_{\mu\nu} = J_{\nu\mu}.
\ee
%%%%%%4%%12%%%10
The functional determinant $det \ U$ is determined by
\be
\frac{1}{det \ U}= \int d[\L^{\rh}]\mbox{\large \boldmath $ \Pi$}_{x,\ld}
\de\left(F_{\ld}(J^\$) -Y_{\ld}\right), \ \ \ \ \ \   F_{\ld}(J^\$)=h_{\ld}^{ \  \mu\nu} J^{\$}_{\mu\nu},
\ee
% %%%%%%%%%%%%23%%%%%%%%31%%%%%%32%%%%11
where the tensor  $J^{\$}_{\mu\nu}$ is given by the $T_4$ gauge 
transformation (3), i.e.,
 \be
 J^{\$}_{\mu\nu}(x) = J_{\mu\nu} + f_{\mu\nu\ld} \Ld^{\ld}, 
 \ee
 %12
 \be
 f_{\mu\nu\ld} =  -  (\p_\ld J_{\mu\nu}) -
J_{\mu\ld} \p_\nu - J_{\ld\nu} \p_\mu. 
\ee
%%%%%%%%13
Moreover, $U_{\rh\ld}$ is given by\cite{8}
\be
U_{\rh\ld}(J) =- \frac{\d F_{\rh}(J)}{\d J_{\mu\nu}} f_{\mu\nu\ld}= - h_{\rh}^{ \  \mu\nu} f_{\mu\nu\ld}.
\ee
%%%%%%%%%%25%%%%%%%%%%33%%%%%%%14
We use $J \equiv J_{\mu\nu}$ to express the gravitational action, the $T_4$ gauge curvature $C^{\mu\nu\a}$, etc. in (9)-(14) and to carry out variations and calculations in Yang-Mills gravity.
Based on (8), we can write the vacuum-to-vacuum amplitude in the form
$$
W = \int d[Y^{\nu}] W(Y^{\ld})
exp\left[i\int d^{4}x\left( \frac{\xi}{2g^{2}}Y^{\mu}Y_{\mu}\right)\right],
$$
%%%26%%%%%%%%%%%%%34%%%%%35%%%%%10.13%15
\be
=\int d[\phi](det \ U) 
exp\left[i\int d^{4}x (L_{\phi} + \frac{\xi}{2g^2}F_\ld F^\ld)\right],
\ee
to within an unimportant multiplicative factor,\cite{8} where $F_{\ld}$ is given by (7).

To generate Green's functions,\cite{17} we introduce external source term $j_{\mu\nu}$ for the $T_4$ gauge fields  in the vacuum-to-vacuum amplitudes (15),
\be
 W(j) =\int d[\phi](det \ U) 
exp\left[i\int d^{4}x (L_{\phi} + \frac{\xi}{2g^2}F_\ld F^\ld + J_{\mu\nu}j^{\mu\nu})\right],
\ee
%16
where $W(j)$, $d[\phi]detU$ and $\int d^4 x L_{\phi}$ are invariant under the infinitesimal gauge transformations.
%%%%%%%

Under the infinitesimal $T_4$ gauge transformations (3) and (4) with the constraint (5), we have
\be
J_{\mu\nu} \to J_{\mu\nu} + \de J_{\mu\nu},   \ \ \ \ \    F_{\nu} \to  F_{\nu} + \de F_{\nu},  \ \ \ \  
W(j) \to W(j),
\ee
%13.10%%%%%%%17
where
\be
\de J_{\mu\nu} =  f_{\mu\nu\ld} \Ld^{\ld}= f_{\mu\nu\ld} [U^{-1} (J)]^{\ld\a} \rho_{\a},
\ee
%%13.11%18
\be
\de F_{\ld}(J)= h_{\ld\mu\nu} f^{\mu\nu\a} \Ld_{\a} = h_{\ld\mu\nu} f^{\mu\nu\a} [U^{-1}(J)]_{\a\b} \rho^{\b} = \rho_{\ld},
\ee
%%13.12%%19 
where we have used (5), while $h_{\ld}^{ \ \mu\nu}$ and $f_{\mu\nu\ld}$ are given in (7) and (13) respectively. 
 From (16) and (17), we have
\be 
\frac{\de W(j)}{\de \rho_{\a}} = \int d[\phi](det \ U) 
exp\left[i\int d^{4}x (L_{\phi} + L_{\xi\zeta} + J_{\mu\nu}j^{\mu\nu})\right]Z^\a = 0,
\ee
%%%%%13.15%%20
where
$$
Z^{\a} =\frac{\de}{\de \rho_{\a}} \int d^4 x \left(  \frac{\xi}{2g^2}F_\a F_\b \e^{\a\b} + J_{\mu\nu}j^{\mu\nu})\right)
$$
\be
=  \frac{\xi}{2g^2}F^\a (J) + j^{\mu\nu}f_{\mu\nu\ld}[U^{-1}(J)]^{\ld\a},
\ee
%%%17%21
 and the gauge fixing terms  $L_{\xi\zeta}$ in the Lagrangian can be written in the form
\be
 L_{\xi\zeta} =  \frac{\xi}{2g^2}F_\ld F^\ld= \frac{\xi}{2g^2} (\p_\mu J^{\mu\ld} - 
\frac{\zeta}{2} \p^\ld J^{\mu}_{\mu})(\p_\nu J^{\nu\rh} - \frac{\zeta}{2} \p^\rh J^{\nu}_{\nu})\e_{\ld\rh}. 
 \ee
%%%%%13.16%%%22
The gauge-fixing terms in (22) involve two arbitrary gauge parameters $\xi$ and $\zeta$ and represent a wide class of gauge conditions.

It follows from (17), (20), (21) and (22), we obtain the space-time gauge identities for the generating functional $W(j)$,  
\be
\left[\left(\frac{\xi}{2g^2}F^\a \left(\d^*\right)\right)_{y}\right.
\left.+\int d^4 z \  j^{\mu\nu}(z)\left(\frac{}{}f_{\mu\nu\ld}\right)_{z}\left[ U^{-1}\left(\d^*\right)\right]_{zy}^{\ld\a}\right]W(j)=0,
\ee
%%%13.18%%23
where $\d^*\equiv (1/i)(\d/\d j)$. For the identities to be convenient for explicit calculations, we have to know $[U^{-1}]W(j)$ in terms of fields.  To accomplish,  let us use (16) with external source and consider\cite{17}
\be
W^{\mu\nu} (j)  =\int d[\phi,V',V] V^{\mu} V'^{\nu}
exp\left[i\int d^{4}x (L_{ef} + j^{\mu\nu}J_{\mu\nu})\right],
\ee
%%%13.19%%24
\be
L_{ef}=L_{\phi}+ L_{\xi\zeta} + V'^{\mu}U_{\mu\nu} V^{\nu},
\ee
%%20%%25
where $det  \ U$ can be described as loops generated by anti-commuting ghost vector fields $V'^{\mu}$ and $V^{\nu}$, 
\be
det \ U = \int d[V',V] exp \left[i \int d^4 x  \ V'^{\mu} U_{\mu\nu}V^{\nu}\right].
\ee 
%26
Such ghost vector-fermions obeyed Fermi-Dirac statics, and were introduced in quantum Einstein gravity by Feynman and DeWitt.\cite{18,19,20}  They are crucial for unitarity and gauge invariance of the theory and  may also be called Feynman-DeWitt ghosts in quantum Yang-Mills gravity.  

The functional (24) involves ghosts $V$ and $V'$ and the external source for gauge fields.  It satisfies\cite{17}
\be
\left[U (\d^*)\right]_{\mu\nu} W^{\nu\ld} (j) = \de^{\ld}_{\mu}W(j).
\ee
%%21%%%27
For the operator $U(\d^*)$, we have
\be
\left[U^{-1}(\d^*)\right]^{\a\mu} \left[U(\d^*)\right]_{\mu\nu}= \de^{\a}_{\nu}, \ \ \ \ \   \d^* \equiv \frac{1}{i}\frac{\de}{\de j}.
\ee
%%%%%%13.22%%%28
It follows from (23)-(28) that
\be
 \int 
exp\left[i\int d^{4}x \left(\frac{}{}L_{ef}+ J_{\mu\nu}(x)j^{\mu\nu}(x)\right)\right] 
\ee
%%%%%%%%%29
$$
\times \left[ \frac{\xi}{g^2}(h_{\mu\nu\ld}J^{\mu\nu})_{y}+\int d^4 z \  V'^{\nu}(y) j^{\mu\nu}(z)\left(\frac{}{}f_{\mu\nu\ld}V^{\ld}\right)_{z}\right] d[\phi,V,V']=0,
$$
where
$$
\left(f_{\mu\nu\ld}V^{\ld}\right)_{z}=\left[-\left(\frac{\p}{\p z^{\ld}}J_{\mu\nu}(z)\right)-J_{\mu\ld}(z)\frac{\p}{\p z^{\nu}}-J_{\ld\nu}(z)\frac{\p}{\p z^{\mu}}\right]V^{\ld}(z), \ \ \  etc.
$$
Thus, we have derived the gravitational space-time gauge identities (29) for the generating functional $W(j)$ in a wide class of gauge conditions specified by (6) and (7).

\section{Space-time gauge identities and a general graviton propagator}

In general, the graviton and ghost propagators in Feynman rules depend on the 
specific form of gauge condition and the gauge 
parameters.  For example, if one chooses the gauge condition specified by (22) involving two arbitrary parameters  $\xi$ and $\zeta$, the graviton
propagator will depend on these two gauge parameters,\cite{21}  
$$
G_{\a\b\rh\s}=\frac{-i}{2 k^{2}}\left(
\left[\e_{\a\b}\e_{\rh\s}- \e_{\rh\a}\e_{\s\b}-\e_{\rh\b}\e_{\s\a}\right]
-\frac{1}{k^{2}}\frac{(2\zeta-2)}{(\zeta-2)}\right.
$$
%%%%%%%%%%12%16%%22%23%%%%30
$$\times \left(k_{\a}k_{\b}\e_{\rh\s}
+k_{\rh}k_{\s}\e_{\a\b}\right)
+\frac{1}{k^{2}}\frac{(\xi-2)}{\xi}(k_{\s}k_{\b}\e_{\rh\a}
+k_{\a}k_{\s}\e_{\rh\b}+ k_{\rh}k_{\b}\e_{\s\a}$$
\be
+k_{\rh}k_{\a}\e_{\s\b})
\left.-\frac{1}{k^{4}}\frac{(2\zeta-2)}{\xi(\zeta-2)}
\left[\frac{4+2\xi-4\xi\zeta}{\zeta-2}-4+2\xi\right]
k_{\a}k_{\b}k_{\rh}k_{\s}\right).
\ee
The $i\epsilon$ prescription for the Feynman propagator is understood.  It reduces to a simpler form
\be
G_{\a\b\rho\s}=\frac{-i}{2 k^{2}}
\left(\frac{}{}\left[\e_{\a\b}\e_{\rh\s}- \e_{\rh\a}\e_{\s\b}-\e_{\rh\b}\e_{\s\a}\right]\right.
\ee
%%%%%%31
$$
+\frac{1}{k^{2}}\frac{(\xi-2)}{\xi}(k_{\s}k_{\b}\e_{\rh\a}
+k_{\a}k_{\s}\e_{\rh\b}+ k_{\rh}k_{\b}\e_{\s\a}
\left.+k_{\rh}k_{\a}\e_{\s\b})\frac{}{}\right)
$$
for $\zeta=1$ and   $\xi$=arbitrary.   
 
The propagator of the ghost vector particle can be derived from the 
effective Lagrangian (25) and the relation (14) for the ghost field,
\be
G^{\mu\nu}=\frac{-i}{k^{2}}
\left(\e^{\mu\nu}-\frac{k^{\mu} 
k^{\nu}}{k^{2}}\frac{(1-\zeta)}{(2-\zeta)}\right),
\ee
%%%%%%%%%%%%%%22%30%%%%%%%%%%%%%%%29%%%32
where $\zeta$ is not equal to 2.  In the special case $\zeta=1$, 
which corresponds to the special gauge condition $\p^\mu J_{\mu\nu} - 
(1/2) \p_\nu J^{\mu}_{\mu} \equiv Y^{\nu}_{1}$.  The effect Lagrangian is given by 
(25) with $\zeta=1$. 

Differentiating the gauge identities (29) with respect  to $j_{\mu\nu}$ and letting all external sources vanish, we obtain the simplest identities,
\be
\xi <T \ (h^{\ld\mu\nu} \phi_{\mu\nu})_{x} \phi^{\a\b} (y)> = - <T  \ V'^{\ld} (x) (f^{\a\b\s} V_{\s})_{y} >,
\ee
%%%33
where $T$ denotes chronological ordering of the fields.
This is a generalization of  Ward-Takahasi-Fradkin identities to the Abelian $T_4$ gauge group because (33) contains Feynman-DeWitt ghosts explicitly.  Thus, the relation (33) may be termed `graviton-ghost gauge identities' of Yang-Mills gravity in flat space-time.

For the lowest order, the identity (33) together with (7) and (13) leads to the identity, 
\be
\xi k^{\b} G_{\a\b\rh\s}(k) - \frac{\xi\zeta}{2}k_{\a} \e^{\mu\nu} G_{\mu\nu\rh\s}(k) = -k_{\s}G_{\a\rh}(k) - k_{\rh} G_{\a\s},
\ee
%%34
where $G_{\a\b\rh\s} (k)$ and $G_{\a\rh}(k)$ are respectively general propagators of the graviton (30) and the Feynman-DeWitt ghost (32).

Direct calculations give the following results, 
\be
A_{\a\rh\s} \equiv \xi k^{\b} G_{\a\b\rh\s}(k) = \frac{-i}{2k^2}\left[- \xi k_{\s}\e_{\rh\a}\frac{\zeta}{\zeta - 2} -2(k_{\s}\e_{\rh\a} + k_{\rh}\e_{\s\a})\right.
\ee
%%%%%35
$$
\left. +\frac{k_{\a}k_{\rh}k_{\s}}{k^2}\left(2\xi -4 +\frac{2\zeta -2}{\zeta -2}[-3 \xi +4 +\frac{4\xi \zeta - 4 - 2\xi}{\zeta -2}]\right)\right],
$$
\be
B_{\a\rh\s} \equiv  -\frac{\xi\zeta}{2} k^{\a}\e^{\mu\nu} G_{\mu\nu\rh\s}(k) = \frac{-i}{2k^2}\left[- \xi k_{\a}\e_{\rh\s}\frac{\zeta}{\zeta - 2}  +\frac{k_{\a}k_{\rh}k_{\s}}{k^2}\right.
\ee
%%%%%36
$$
\left. \times \left(-2\xi\zeta +4\zeta +\frac{\zeta(2\zeta -2)}{\zeta -2}\left[-3 \xi -2 - \frac{2\xi \zeta - 2 - \xi}{\zeta -2}\right]\right)\right],
$$
The left-hand-side of the identities (34) are the sum of (35) and (36),
\be
A_{\a\rh\s} + B_{\a\rh\s} =\frac{-i}{k^2}\left[-(k_{\s}\e_{\rh\a} + k_{\rh}\e_{\s\a}) +\frac{2\zeta-2}{\zeta-2}\left(\frac{k_{\a}k_{\rh}k_{\s}}{k^2}\right)\right].
\ee
%%%%%%%%37
Using (32), the right-hand-side of (34) gives the result,
\be
 -k_{\s}G_{\a\rh}(k) - k_{\rh} G_{\a\s} = \frac{-i}{k^2}\left[-k_{\s}\e_{\a\rh} - k_{\rh}\e_{\a\s} + \frac{2k_{\s}k_{\a}k_{\rh}(1-\zeta)}{k^2 (2-\zeta)}\right].
\ee
The results (37) and (38) imply that the space-time gauge identities (34) for a general gauge condition are substantiated by explicit calculations with a very general graviton propagator involving two arbitrary gauge parameters $\xi$ and $\zeta$.  The calculation of identity for the next order with one-loop correction is formidable and needs the help of symbolic computing of a computer.\cite{22}

 \section{Discussions}
 
The graviton-ghost gauge identities (33) are not unique.  There is another form of space-time gauge identities, in which the Feynman-DeWitt ghosts do not show up explicitly, in contrast to (33). As in the usual gauge theories,\cite{23}  the vacuum-to-vacuum amplitude $W$ in (15) can be written in the form
\be
W=\int d[Y^\nu] W(Y^{\ld}) exp\left[i \int \frac{\xi}{2 g^2}(Y^\mu - B^{\mu})(Y_\mu - B_\mu)\right] 
\ee
%%%39
$$
=\int d[\phi] det U exp\left[i \int d^4 x \left( L_\phi + \frac{\xi}{2g^2} (F^\mu - B^\mu) (F_\mu - B_\mu) \right)\right],
$$
where $F^{\ld}=h_{\ld}^{ \  \mu\nu} J_{\mu\nu} $.  The expression (39) is  independent of the arbitrary vector funciton of space-time $B_{\mu}$.  One can expand $W$ in powers of $B_{\mu}(x)$.   All coefficients of powers of $B_{\mu}$ must vanish, except the zeroth order.  Using the relation $\int B^\mu (x) B_{\mu}(x) d^4 x = \int B^\mu (x) B^{\nu}(y)\e_{\mu\nu} \de^4 (x-y) d^4 y \  d^4 x $, the coefficient of the power $B_{\a}(x) B_{\b} (y)$ leads to the following space-time gauge identities:
\be
i \xi \langle T ((\p_{\mu} \phi^{\mu\a} - \frac{\zeta}{2} \p^{\a} \phi^{\mu} _{\mu})_{x}(\p_{\nu} \phi^{\nu\b} - \frac{\zeta}{2} \p^{\b} \phi^{\nu} _{\nu})_{y}) \rangle = - \e^{\a\b} \de^4 (x-y),
\ee
%%%%%%%%%40
where $(\p_{\mu} \phi^{\mu\a})_{y}=\p \phi^{\mu\a} (y)/ \p y^{\mu}$, etc.  The gauge identities (40) are not independent 
of (33).\footnote{In fact, (40) can also be derived from (23).  The derivation for the special case $\zeta=0$ is easier to carry out.}   The  result  (40) may be termed `graviton gauge identities' because the Feynman-DeWitt ghosts do not show up explicitly, in contrast to (33).  It is more convenient to use (40) to illustrate the relations of `renormalization' constants of the graviton propagator and the gauge parameter $\xi$.  (See Appendix.)

From the explicit calculations of the space-time gauge identities (34)-(38), one can see cancellations of a large number of  terms and feel the power of space-time gauge symmetry.  The situation resembles that in non-Abelian gauge field theory:  When one calculates the imaginary part of the amplitudes coming from the interactions of, say, the unphysical component of a vector boson and scalar ghost to verify unitarity of physical amplitudes, we also see very many cancellations of unphysical amplitudes involving gauge parameters, so that unitarity of the S matrix is preserve.  From these calculations, we can feel the power of gauge symmetry of internal groups.\cite{24}
  
 One may get the impression that, say, quantum chromodynamics has ghost particles because its gauge groups $[SU_3] _{color}$ is non-Abelian.  Moreover, quantum electrodynamics (QED) with the Abelian group $U_1$ does not have ghosts.  But this is true if and only if one imposes a linear gauge condition in QED.  To illustrate the implications of gauge symmetry  in a generalized Ward-Takahasi-Fradkin (WTF) identities, we consider QED with a class of non-linear gauge condition\cite{1} 
\be
F(A) \equiv \p_{\mu}A^{\mu} - \b' A_{\mu}A^{\mu} = a(x),   \ \ \  \b' \ne 0,
\ee
%%39%%%41
where $a(x)$ is a real function independent of $A_{\mu}$ and $\Ld (x)$.  

 In the absence of the electron, the gauge invariant Lagrangian for the photons implies that photons do not couple with themselves and are free particles.  However, when one choose a non-linear gauge condition (41), there will be new 3-vertex and 4-vertex  self-coupling  of photons.  These new self-coupling of photon will upset gauge invariance and  unitarity of QED.  The generalized WTF identities assure that the extra unwanted amplitudes produced by these new vertices will be cancelled by the interactions of the electromagnetic ghosts, so that gauge invariance and unitarity of physical amplitudes are preserved. 

The generating functional of Green's function in the non-linear gauge can be written as\cite{1}
\be
W_{\a}(\e, \overline{\e}, \e_{\mu})=\int d[\Om] (det  \ M) exp[i \int (L + L_s -\frac{1}{2\a}F^{2})d^4 x
\ee
%%%40%%%42
\be
L=-\frac{1}{4} F_{\mu\nu}F^{\mu\nu}+ L_{\psi} ,  \ \ \   L_{s} = \e_{\mu}A^{\mu} + \overline{\e}\psi + \overline {\psi} \e, \ \ \  [\Om] \equiv[A_{\mu},\overline{\psi}, \psi]
\ee
%41%%%43
\be
M=(\p_{\mu} - 2\b' A_{\mu})\p^{\mu}, \ \ \ \ \ \    detM = \int  d[c',c] exp\left(i\int d^4 x   \ c' M c \right).
\ee
%%%42%%%%44
where $F_{\mu\nu}=\p_{\mu} A_{\nu} - \p_{\nu} A_{\mu}$ and $L_{\psi}$ is the usual fermion Lagrangian in QED.  The generalized WTF identities was obtained in the form
$$
\left[ -\frac{1}{\a}\left(\p_{\mu} \frac{\d}{i\d \e_{\mu}} + \b'\frac{\d^2}{\d \e^{\mu} \d \e_{\mu}} \right) \right.
\left. +\e_{\mu} \p^{\mu} M^{-1} \right.
$$
\be
-e\frac{\d}{\d \e} \e M^{-1} - \left. e\overline{\e}\frac{\d}{\d \overline{\e}} M^{-1} \right] W_{\a}(\e, \overline{\e}, \e_{\mu}) = 0.
\ee
%%%43%%%%45
Here we have used the gauge transformation $A'_{\mu}=A_{\mu} +\p_{\mu}\Ld(x)$ with the constraint $\Ld(x) = M^{-1}\rh(x)$, where $\rh(x)$ is an arbitrary infinitesimal real function independent of $A_{\mu}$.  Following similar steps from (23)  to (29), we can express the generalized WTF identities (45) in the form
\be
 \int 
exp\left[i\int d^{4}x \left(\frac{}{}L_{eff}+ A_{\mu}\e^{\mu}+ \overline{\e}\psi + \overline {\psi} \e\right)\right] 
\ee
%%%%%%%%%44%%%46
$$
\times \left[ \frac{1}{\a}[(\p_{\mu} - \b' A_{\mu})A^{\mu}]_{y}+\int d^4 z \  c'(y) \e^{\mu}(z)[\p_{\mu}c]_{z}\right] d[\Om,c,c']=0,
$$
where
$$
L_{eff}=L+ \frac{1}{2\a}(\p_{\mu}A^{\mu} - \b' A_{\mu}A^{\mu})^{2} + c'Mc,    \ \ \   [\p_{\mu}c]_{z}=\frac{\p}{\p z^{\mu}}c(z),  \ \  etc.  
$$
%%20%%45%%%%%47
Thus, we have derived the generalized WTF identities involving electromagnetic ghosts $c'(x)$ and $c(x)$  for the generating functional $W_{\a}$ in the non-linear gauge specified by (41).

To the lowest order, the simplest identities obtained from (46)  are given by
\be
\frac{1}{\a} k^{\mu}G_{\mu\nu}(k) + k_{\nu}G(k) = 0.
\ee
%%%%%%46%%%45%%%%48
which corresponds to (34).  The propagators for the photon and the electromagnetic ghost are respectively given by $G_{\mu\nu}(k)$ and $G(k)$, \cite{1}
\be
G_{\mu\nu}(k)=\frac{-i}{k^2}\left(\e_{\mu\nu} - (1-\a)\frac{k_{\mu}k_{\nu}}{k^2}\right), \ \ \ \ \ \  G(k)=\frac{i}{k^2},
\ee
%%%%%%%47%%%46%%%49
where the $i\epsilon$ prescription for the propagator is understood. One can verify that the identity (48) is indeed satisfied.

The underlying flat space-time for Yang-Mills gravity and its unification with other interactions\cite{12,13} is probably fundamental to all known interactions in nature and critical for their unification.  Such a space-time framework can accommodate all the essentials requirements for physics such as all conservation laws, all physically realizable frames of reference (both inertial and non-inertial)\cite{14} and the quantization of all physical (and `ghost') fields associated with internal and external gauge symmetries.  Thus, it may be termed a `taiji symmetry framework' since the word `taiji' denotes, in ancient Chinese thought, the ultimate principles or the conditions that existed before the creation of the world.  One of problems with employing a more general framework based on curved space-time with general coordinate invariance is that Noether's (unsung) theorem II implies that because there is a continuously infinite number of generators in such a symmetry framework and thus there is no conservation law for energy.\cite{25}

For a general discussion to all orders in perturbative Yang-Mills gravity, the space-time gauge identities are essential for reducing the number of independent renormalization constants.  Although Yang-Mills gravity is not renormalizable in the usual sense of power-counting, the structure of its interaction has  similarities to  usual renormalizable gauge theories.  For example, the maximum number of graviton  self-coupling in a vertex is 4, as shown in the Lagrangian $L_{\phi}$ (9) in Yang-Mills gravity.  
This maximum 4-vertex for graviton interaction together with the T(4) gauge symmetry in Yang-Mills gravity and the powerful dimensional regularization
could shed light on quantum gravity.   
 \bigskip

\noindent
{\bf Acknowledgements}
\bigskip

The author would like to thank S. H. Kim for his help to confirm some results using symbolic computing.   The work was supported in part by  Jing Shin
Research Fund and Prof. Leung Memorial Fund of the UMass Dartmouth Foundation.

\bigskip

\bigskip

\bigskip

\bigskip
\noindent
{\bf Appendix}
\bigskip

\noindent
{\bf  Relations of `renormalization' constants for propagator and $\xi$}
\bigskip

In gauge theory with $SU_2$ symmetry, the Slavnov-Taylor identities imply that the 
counter-term of the mass renormalization vanishes.\cite{6,7}  Nevertheless, the space-time gauge identities in Yang-Mills gravity do not have the exact analog.   In general, the situation of relations for renormalization constants is much more complicated in Yang-Mills gravity because by power counting the  theory is not renormalizable.    One must include a new counter-term for mass renormalization of graviton in Yang-Mills gravity in flat space-time.  The situation is similar to the case of $\g_5$ meson-nucleon interaction  theory, where one must include a new counter term of quartic meson coupling for renormalization.\cite{26}

Let us demonstrate that space-time gauge identities can lead to a set of relations useful to `renormalization' of graviton propagator and gauge parameters $\xi$.  Since Yang-Mills gravity is not renormalizable in the usual sense of power counting, we may restrict ourselves to use `renormalization' of the divergent quantities at the one-loop level.  Fortunately, the dimensional regularization allows us to discuss sensibly renormalization problem in Yang-Mills gravity.\footnote{The powerful dimensional regularization and space-time translational gauge symmetry tempt one to the conjecture that Yang-Mills gravity in flat space-time may be renormalizable.  At least, all divergent terms can be removed by counter terms in the Lagrangian.}  For simplicity, let us set $\zeta =1$ and leave $\xi$ arbitrary in the graviton propagator (30) and the graviton gauge identity (40).  Similar to the situation in non-Abelian gauge theory,\cite{17} one can parametrize the graviton propagator (31)  as follows:
\be
G^{\a\b\rh\s}(k) = \frac{-i}{2}\left[ \frac{1}{a}\left(\e^{\a\b}\e^{\rh\s} - \e^{\a\rh} \e^{\b\s} - \e^{\a\s}\e^{\b\rh}\right)\right.
\ee
%%%%%%49
$$
\left.+\frac{c}{ab} (k_{\s}k_{\b}\e_{\rh\a}
+k_{\a}k_{\s}\e_{\rh\b}+ k_{\rh}k_{\b}\e_{\s\a}
+k_{\rh}k_{\a}\e_{\s\b})\right],
$$
where a, b and c may  be functions of $k^2$.  These unknown functions may involve `renormalization' constants and are to be determined by the graviton gauge identities (40).  With the help of Fourier transform, the four terms on the LHS of (40) can be expressed in terms of the graviton propagator in momentum space:
\be
\langle T( (\p_{\a} \phi^{\a\b})_{x} ( \p_{\rh} \phi^{\rh\s})_{y}) \rangle  \to (ik_{\a})(-i k_{\rh} )G^{\a\b\rh\s}(k) 
\ee
%%%%%50
$$
=\frac{-i}{2}\left[\frac{-k^2 \eta^{\b\s}}{a} +\frac{c}{ab}(3 k^2 k^\s k^\b + k^4 \e^{\s\b})\right]
$$

\be
-\frac{1}{2}\langle T( (\p^{\b} \phi^{\mu} _{\mu})_{x}(\p_{\rh} \phi^{\rh\s} )_{y}
+ (\p_{\mu} \phi^{\mu\b})_{x}(\p^{\s} \phi^{\nu}_{\nu} )_{y}) \rangle \ \  \to  \ \
\ee
%%%%%%%%%%%%51
$$
=\frac{-i}{2}\left[\frac{2 k^{\b} k^{\s}}{a} +\frac{c}{ab}(4 k^2 k^\s k^\b) \right]
$$
\be
\frac{1}{4} \langle T( (\p^{\b} \phi^{\mu} _{\mu})_{x}(\p^{\s} \phi^{\nu}_{\nu})_{y})\rangle \ \ \to  \ \  =\frac{-i}{8}\left[\frac{8 k^{\b} k^{\s}}{a} +\frac{c}{ab}( 4 k^2 k^\s k^\b )\right].
\ee
%%%%%%%%52
Using (50)-(52) and $- \e^{\s\b} \de^4 (x-y)  \to \  -\e^{\s\b}$, the graviton gauge identities (40) gives
\be
\frac{p^2 \e^{\a\b}}{a}\left(\frac{1}{2} - \frac{c p^2}{2b} - \frac{a}{\xi p^2}\right) +\frac{p^\a p^\b}{a} Q = 0,
\ee
%%%%%%53
where
$$
Q= \left(\frac{-3c p^2}{2b} +1 +2\frac{c p^2}{b} -1 -\frac{c p^2}{2b}\right) = 0.
$$
One can `renormalize' massless fields and parameters as usual so that
\be
a(p^2)= Z_{3}^{-1} p^2,  \ \ \ \    b(p^2)= {Z'}_{3}^{-1} p^2, 
\ee
%%%%%%%%54
for $p^2 \to 0$, where $Z'_{3}$ may not be the same as $Z_{3}$ in general.  It follows from (53) and (54) that
\be
  c {Z'}_{3}  = \frac{\xi_{r} - 2}{\xi_{r}},      \ \ \ \ \ \    \xi_{r} = Z_3 \xi, 
\ee
%%%%%%%%55
where $Z_3$ is the `renormalization' constant for the gravitational wave function $\phi_{\mu\nu}$  and  $\xi_{r}$ is the renormalized gauge parameter.  The result (55) is consistent with the form (31) of the graviton propagator.

%% Bibliography
%%%%%%%%%%%%%%%%%%%
%\newpage
%\section*{References}

\bibliographystyle{unsrt}

\begin{thebibliography}{99}
\bibitem{21}There are sign errors in $G_{\a\b\rh\s}$ and a missing factor (1/2) in eq. (23) of the previous paper in ref.8.  They are corrected in eq. (30) here, which is shown to be consistent with the translational gauge identities in equations (34)-(38).
\end{thebibliography}

\end{document}